Title: Remarks on the instability of black Dp-branes
Authors: J. X. Lu and Shibaji Roy
Comments: 12 pages, one figure, v3: added free energy computation, version to
appear in PLB

We show that for black D$p$-branes having charge $Q$ and Hawking temperature
$T$, the product $QT^{7-p}$ is bounded from above for $p\leq 5$
and is unbounded for $p=6$. While the maximum occurs at
some finite value of a parameter for $p \leq 4$, it occurs at infinity
of the parameter for $p=5$.
As a consequence, for fixed charge, there are two black D$p$-branes
(for $p\leq 4$) at any given temperature less than its maximum value, and
when the temperature is maximum there is one black D$p$-brane.
For $p=5$, there is only one black D5-brane at a given temperature less
than its maximum value, whereas, for $p=6$, since
there is no bound for the temperature, there is always a black D6-brane
solution at a given temperature.  Of the two black D$p$-branes (for
$p\leq 4$), one is large which is shown to be thermodynamically unstable and
the other is small which is stable. But for $p=5,6$, the black D$p$-branes
are always thermodynamically unstable. The stable, small black D$p$-brane,
however, under certain conditions, can become unstable quantum mechanically
and decay either to a BPS D$p$-brane or to a Kaluza-Klein
``bubble of nothing'' through closed string tachyon condensation. The small
D5, D6 branes, although classically unstable, have the same fate under similar
conditions.

\documentclass[12pt]{article}
\input epsf.sty

\newcommand{\bea}{\begin{eqnarray}}
\newcommand{\eea}{\end{eqnarray}}
\newcommand{\be}{\begin{equation}}
\newcommand{\ee}{\end{equation}}
\newcommand{\vs}[1]{\vspace{#1 mm}}

\newcommand{\dsl}{\pa \kern-0.5em /}

\newcommand{\pa}{\partial}

\newcommand{\nn}{\nonumber\\}

\newcommand{\eqn}[1]{(\ref{#1})}

\begin{document}
\topmargin 0pt
\oddsidemargin 0mm

\begin{flushright}

USTC-ICTS-09-16\\




\end{flushright}

\vspace{2mm}

\begin{center}

{\Large \bf Remarks on the instability of black D$p$-branes}
\vs{6}

{\large J. X. Lu$^a$\footnote{E-mail: jxlu@ustc.edu.cn} and Shibaji
Roy$^b$\footnote{E-mail: shibaji.roy@saha.ac.in}}

 \vspace{4mm}

{\em

 $^a$ Interdisciplinary Center for Theoretical Study\\

 University of Science and Technology of China, Hefei, Anhui
 230026, China\\

\vs{4}

 $^b$ Saha Institute of Nuclear Physics,
 1/AF Bidhannagar, Calcutta-700 064, India}

\end{center}

\vs{10}

\begin{abstract}

We show that for black D$p$-branes having charge $Q$ and Hawking temperature
$T$, the product $QT^{7-p}$ is bounded from above for $p\leq 5$
and is unbounded for $p=6$. While the maximum occurs at
some finite value of a parameter for $p \leq 4$, it occurs at infinity
of the parameter for $p=5$.
As a consequence, for fixed charge, there are two black D$p$-branes
(for $p\leq 4$) at any given temperature less than its maximum value, and
when the temperature is maximum there is one black D$p$-brane.
For $p=5$, there is only one black D5-brane at a given temperature less
than its maximum value, whereas, for $p=6$, since
there is no bound for the temperature, there is always a black D6-brane
solution at a given temperature.  Of the two black D$p$-branes (for
$p\leq 4$), one is large which is shown to be thermodynamically unstable and
the other is small which is stable. But for $p=5,6$, the black D$p$-branes
are always thermodynamically unstable. The stable, small black D$p$-brane,
however, under certain conditions, can become unstable quantum mechanically
and decay either to a BPS D$p$-brane or to a Kaluza-Klein
``bubble of nothing'' through closed string tachyon condensation. The small
D5, D6 branes, although classically unstable, have the same fate under similar
conditions.

\end{abstract}

\newpage


Black D$p$-branes are the low energy solutions of string theory
\cite{Horowitz:1991cd, Duff:1993ye} and can
be considered as the higher dimensional analog (without the matter fields)
of black holes in general
theory of relativity. It is true that the black holes in four dimensions are
classically stable, but, it is not so for their higher dimensional
cousins. This was first shown by Gregory and Laflamme \cite{Gregory:1993vy}
for the neutral black
string solution in space-time dimensions $d=5$. They showed that this
space-time under linearized perturbation suffers from a long wavelength
instability. This feature is quite generic and remains true even when
one introduces charge to the black string solution \cite{Gregory:1994bj}
except when the extremal point is reached \cite{Gregory:1994tw}. However,
the stability
can be restored when the string direction is compactified and the size
of the compact
direction is made smaller than the horizon radius. The end point of the
Gregory-Laflamme instability is believed to be the non-uniform or
inhomogeneous string \cite{Horowitz:2001cz, Horowitz:2002ym}
as the full dynamics of the transition is difficult to
study in the linearized perturbation and remains somewhat speculative.
Later it has been shown by Gubser and Mitra \cite{Gubser:2000ec,
Gubser:2000mm} that the
(uncompactified) charged black string will suffer from Gregory-Laflamme
instability exactly when their specific heat is negative known as the
correlated stability conjecture. Similar analysis also holds true for the
black and neutral as well as charged $p$-brane solutions in higher dimensions
with $p>1$ \cite{Reall:2001ag} (also see \cite{Harmark:2007md}
for a recent review).

In this letter we will specifically consider the black D$p$-brane
solutions of type IIA/B string theories\footnote{The black string
solution in $d=5$ considered by Gregory and Laflamme
\cite{Gregory:1993vy} is nothing but the double dimensionally
reduced black D6-brane solution of type IIA string theory. Whereas
the strings in $d>5$ dimensions are the double dimensionally reduced
D$p$-branes (where $p=11-d$) in string theory.}. We will show how
some simple observations can make the classical stability analysis
of these solutions much easier. In particular, we point out that the
charge $Q$ and the Hawking temperature $T$ of the black D$p$-branes
\cite{Horowitz:1991cd, Gubser:1996de,
  Klebanov:1996un} are correlated in such a way that the
product $QT^{7-p}$ has an upper bound for $p\leq 5$.
However, no such bound exists for
D6-brane. This implies that for fixed charge, the temperature of the
D$p$-branes (for $p\leq 5$) has an upper bound, but not for
D6-brane\footnote{This remains also
true if we exchange the role between the charge and the
temperature.}. The temperature becomes maximum at a finite value of a
parameter ($\theta$) related to the ADM mass and the charge of the
solution only for $p\leq 4$, but for $p=5$, the temperature becomes
maximum when $\theta$ becomes infinite. For $p=6$, there is no upper
limit for the temperature. As a consequence we will show that when
$p \leq 4$, there are two black D$p$-brane (one small and the other
large) solutions at any given temperature less than the maximum
temperature and one black D$p$-brane solution when the temperature
is the maximum. For $p=5$ there is one black D5-brane solution for
the temperature less than its maximum value and for D6-brane, as
there is no maximum temperature, there is always a solution for any
given temperature. Using thermodynamics of the black D$p$-brane we
calculate their specific heat \cite{Harmark:1999xt, Gubser:2000mm}
and found that the specific heat is always negative for the large
black D$p$-brane with $p\leq 4$ and it diverges for the black brane
whose temperature is the maximum, signalling a possible critical
point of second order phase transition. But for the small D$p$-brane
with $p\leq 4$, the specific heat is positive. So, based on the
correlated stability conjecture mentioned earlier, the small
D$p$-brane is classically stable for $0 < p \leq 4$ whereas
the large one is classically unstable\footnote{In order to relate the
classical instability to local thermodynamic instability, the gravitational
system must be of infinite extent and this excludes the case $p=0$.}.
For D5 and D6, the specific heat is
always negative and so, they are always classically unstable. The
unstable branes would of course decay to more stable and
entropically more favorable solutions, the dynamics of which may be
difficult to understand in the linearized approximation. But one can
argue from the earlier studies that the large black D$p$-branes,
would either decay into small black D$p$-branes or
to non-uniform (inhomogeneous) D$p$-branes and D5, D6 branes would
decay only into non-uniform (inhomogeneous) branes
\cite{Horowitz:2002ym}.

Thus far we have only discussed about the classical stability of the
black D$p$-branes and mentioned that the small black D$p$-branes are
classically stable for $0 < p \leq 4$. The large black D$p$-branes
(for $0 < p \leq 4$) as well as the black D5 and D6 branes are
classically unstable. However, under certain conditions the
classically stable small black D$p$-branes could become quantum
mechanically unstable. This can happen when we reduce the
temperature of the black D$p$-brane keeping the charge fixed (or
when we reduce the charge while keeping the temperature fixed). By
this, the parameter $\theta$ will become very large and the size of
the black D$p$-brane will get reduced. At this point the issue of
quantum instability for the black D$p$-brane arises. As a result, the
black D$p$-brane will start to Hawking radiate and will smoothly
make a transition to the stable BPS D$p$-brane. This can happen for
all $p\leq 4$. However, when $p>0$\footnote{The $p = 0$ case, though
not quite relevant in the present discussion, can also be addressed
as in \cite{Green:2006nv}.}, there can be another kind of quantum
mechanical instability if one of the brane directions is compact.
When the fermions satisfy periodic boundary condition along the
compact direction, the black brane will again make a transition to
BPS D$p$-brane. On the other hand, when the fermions satisfy
antiperiodic boundary condition, then a fundamental string wound
around the compact direction can become tachyonic \cite{Rohm:1983aq}
and in that case the black D$p$-brane can make a transition to the
static Kaluza-Klein (KK) ``bubble of nothing'' (BON)
\cite{Witten:1981gj} by closed string tachyon condensation
\cite{Horowitz:2005vp, Lu:2007bu}. As we have mentioned, D5 and D6
branes are always classically unstable. However, for very large
$\theta$ (i.e., for small black branes), both D5 and D6 brane can
either decay to non-uniform (or inhomogeneous) branes or they can
smoothly make a transition to BPS branes. However, when one of the
directions of the brane is compact, then the black brane can make a
transition to BPS brane or dynamical KK BON, depending on whether
the fermions in the theory satisfy periodic or antiperiodic boundary
condition along the compact direction, respectively.

The electrically charged black D$p$-brane solution in low energy
type IIA/B string theory is given as \cite{Horowitz:1991cd},
\bea\label{blackbranesoln} ds^2 &=& H^{-\frac{1}{2}} \left(-f dt^2 +
\sum_{i=1}^p(dx^i)^2\right) + H^{\frac{1}{2}}\left(\frac{dr^2}{f} +
r^2 d\Omega_{8-p}^2\right)\nn e^{2\phi} &=& g_s^2 H^{\frac{3-p}{2}},
\quad A_{012\ldots p}\,=\, \frac{1}{g_s} \left(H^{-1} -1\right)
\coth\theta \eea where \be\label{functions} H = 1 +
\frac{r_0^{7-p}\sinh^2\theta}{r^{7-p}}, \quad {\rm and} \quad f = 1
- \frac{r_0^{7-p}}{r^{7-p}} \ee are two harmonic functions and $r_0$
and $\theta$ are two parameters\footnote{Without any loss of
generality, we will assume that the branes are positively charged
and so, $\theta \geq 0$.} characterizing the solution related to the
ADM mass and the charge of the black D$p$-brane. Note that the
metric is written in the string frame and $g_s$ is the string
coupling constant. The black brane has a horizon at $r=r_0$. The
electric charge associated with the black D$p$-brane can be obtained
as, \be\label{charge} Q = \frac{1}{2\kappa^2}\int_{\Omega_{8-p}}
\ast F_{[p+2]} = \frac{(7-p)\Omega_{8-p}}{2\kappa^2 g_s}
r_0^{7-p}\sinh\theta\cosh\theta \ee where $2\kappa^2 = (2\pi)^7
\alpha'^4$, with $2\pi \alpha' = 1/T$, i.e., the inverse of string
tension, and $\Omega_n = 2\pi^{(n+1)/2}/\Gamma((n+1)/2)$ is the
volume of the $n$-dimensional unit sphere. $\ast F_{[p+2]}$ is the
Hodge dual of the field-strength associated with the gauge field
$A_{012\ldots p}$. The Hawking temperature of the black D$p$-brane
can be obtained from the metric in \eqn{blackbranesoln} as usual by
first Euclideanizing the time coordinate and then calculating the
inverse of the periodicity of the Euclidean time direction to avoid
the conical singularity. The temperature has the form,
\be\label{temp} T = \frac{7-p}{4\pi r_0\cosh\theta} \ee Eliminating
$r_0$ from \eqn{charge} and \eqn{temp} we have \be\label{ctheta}
QT^{7-p} = \frac{\Omega_{8-p}}{2\kappa^2
g_s}\frac{(7-p)^{8-p}}{(4\pi)^{7-p}}
\frac{\sinh\theta}{\cosh^{6-p}\theta} =
\frac{\Omega_{8-p}}{2\kappa^2
  g_s}\frac{(7-p)^{8-p}}{(4\pi)^{7-p}} C(\theta)
\ee
From \eqn{ctheta} we notice that $C(\theta) \propto QT^{7-p}$ is positive in
general and vanishes for both
$\theta \to 0$ and $\theta \to \infty$ for $p\leq 4$. But for $p=5,6$,
$C(\theta)$ vanishes only for $\theta \to 0$. $C(\theta)$ becomes constant
as $\theta \to \infty$ for $p=5$, while $C(\theta)$ increases monotonically
with $\theta$ for $p=6$. It can be checked from the form of $C(\theta)$ in
\eqn{ctheta} that while $C(\theta)$ does not have a maximum for $p=6$, there
is a unique maximum of $C(\theta)$ for $p\leq 5$ at $\sinh^2\theta = 1/(5-p)$.
So, for $p \leq 4$, the maximum occurs at a finite $\theta$, but for $p=5$
the maximum occurs when $\theta \to \infty$. The maximum value or
the upper bound of $QT^{7-p}$ can be obtained as,
\be\label{maxvalue}
(QT^{7-p})_{\rm max} = \frac{\Omega_{8-p}}{2\kappa^2
  g_s}\frac{(7-p)^{8-p}}{(4\pi)^{7-p}} C(\theta)_{\rm max} =
\frac{\Omega_{8-p}}{2\kappa^2 g_s}\frac{(7-p)^{8-p} (5-p)^{\frac{5-p}{2}}}
{(4\pi)^{7-p} (6-p)^{\frac{6-p}{2}}}
\ee
So for any given product $QT^{7-p} \propto C(\theta) < C_{\rm max}$, we will
have two solutions for $\theta$ (the smaller $\theta$ is called $\theta_s$
and the larger $\theta$ is called $\theta_{\ell}$) for $p \leq 4$ and one
solution for $p=5,6$. These features of the black D$p$ branes for various
values of $p$ are depicted in Figure 1.

\begin{figure}[!ht]
\leavevmode
\begin{center}
\epsfysize=5cm
\epsfbox{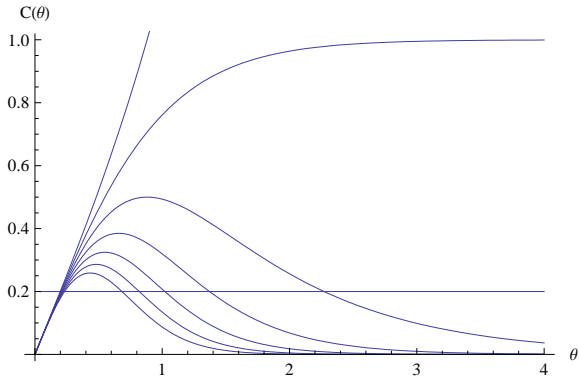}
\end{center}
\caption{The plot shows how $C(\theta)$ varies with $\theta$ in some
suitable units for various black D$p$-branes. The lowest curve is
for $p=0$ and then $p=1$, $p=2$, and so on upto $p=6$. The
horizontal line represents a particular value of $C(\theta)$ and
shows that it cuts the curves at two points for $0\leq p\leq 4$
corresponding to large (the left one) and small (the right one)
black D$p$-branes. For $p=5,6$, the line cuts the curve once for
either case.}\label{figure1}
\end{figure}

Thus for a fixed charge $Q$, as we lower the temperature of the
black D$p$ brane from its maximum value $T_{\rm max}$, at a given
temperature (denoted by the straight horizontal line in Figure 1)
there are two black D$p$-branes (for $p\leq 4$) corresponding to
$\theta_s$ and $\theta_{\ell}$ but there is only one black D5 or D6
brane. Note that when $Q$ is fixed the size of the black brane
corresponding to $\theta_s$ and $\theta_{\ell}$ can be obtained from
\eqn{charge} and this shows that smaller $\theta$ (or $\theta_s$)
corresponds to larger black-brane and larger $\theta$ (or
$\theta_{\ell}$) corresponds to smaller black brane.

Now we try to understand the classical stability of the black D$p$-branes
from thermodynamics. For this purpose we need the expression for the ADM mass
\cite{Lu:1993vt} and the entropy \cite{Klebanov:1996un} of the black
D$p$-brane which
can be calculated from the metric in \eqn{blackbranesoln} and they have the
forms,
\be\label{massentropy}
M = \frac{(8-p)\Omega_{8-p}V_p r_0^{7-p}}{2\kappa^2
  g_s^2}\left(1+\frac{7-p}{8-p} \sinh^2\theta\right),\quad
S = \frac{4\pi\Omega_{8-p}V_p}{2\kappa^2 g_s^2}r_0^{8-p}\cosh\theta
\ee From \eqn{charge}, \eqn{temp} and \eqn{massentropy}, it is clear
that the ADM mass satisfies the Smarr relation (see, for example,
\cite{Harmark:1999xt}), \be\label{smarr} (7-p)M = (8-p) TS + (7-p)
\mu Q \ee where $\mu = \tanh\theta/g_s$ is the chemical potential.
Now using the thermodynamical relation, \be\label{thermo} dM =
TdS+\mu dQ \ee we obtain \be\label{spheat} C_Q \equiv
\left(\frac{\partial M}{\partial T}\right)_Q = T\left(\frac{\partial
S} {\partial T}\right)_Q = \frac{4\pi\Omega_{8-p} V_p \cosh\theta
r_0^{8-p}} {2\kappa^2 g_s^2}\frac{[(9-p)\sinh^2\theta +
(8-p)]}{[(5-p)\sinh^2\theta -1]} \ee where $C_Q$ is the specific
heat of the black D$p$-brane and in writing the second expression in
\eqn{spheat} we have made use of $Q$ in \eqn{charge}, $T$ in
\eqn{temp} and $S$ in \eqn{massentropy}. Now it is clear that the
sign of the specific heat will depend on the factor
$(5-p)\sinh^2\theta -1$ in the denominator of \eqn{spheat}, since
everything else is positive. It should be noted that the factor
$(5-p)\sinh^2\theta -1$ is always negative for $p=5,6$ and so the
black D5 and D6 branes are classically unstable. For $p\leq 4$ the
sign of this factor depends on the value of the parameter $\theta$.
However we find that in general $((5-p)\sinh^2\theta -1)/\cosh^{7-p}
\theta = -(dC(\theta)/d\theta)$, where $C(\theta)$ is defined in
\eqn{ctheta}. So, when the slope of the curve $C(\theta)$ vs
$\theta$ is positive the specific heat is negative and when it is
negative the specific heat is positive. In particular, when
$\theta=0$ (i.e., for chargeless black D$p$-brane), the specific
heat is always negative and the corresponding black D$p$-brane is
classically unstable for $p > 0$. This is Gregory-Laflamme or
Gubser-Mitra instability. By looking at Figure 1, it is clear that
for $p \leq 4$, the slope $dC(\theta)/d\theta$ remains positive
from $\theta=0$ upto the point where $C(\theta)$ becomes maximum
($C_{\rm max}$) and so, the black brane will remain unstable
throughout this region. Then at the maximum point the slope vanishes
and the specific heat diverges. After that the slope becomes
negative from $\theta$ corresponding to $C_{\rm max}$ upto $\theta
\to \infty$. In this entire region the specific heat remains
positive and the black brane is classically stable. In particular,
when $\theta \to \infty$ and also $r_0 \to 0$ in such a way, that
$Q$ remains unchanged, we get back BPS D$p$-brane solution from
\eqn{blackbranesoln} which is obviously stable. This can happen also
for D5 and D6 branes. Since the smaller $\theta$ is for the larger
black D$p$-brane (for $0< p\leq 4$) which is unstable while the
larger $\theta$ is for smaller one which is stable due to (3) for
fixed charge,  we therefore find that large black branes are
unstable and small ones are stable classically for $0 < p \le 4$.

The unstable black branes will of course decay to a more stable or
entropically more favorable solution. As we have mentioned in the
introduction the unstable large black D$p$-branes (for $p \leq 4$)
can decay either into the stable small black D$p$-branes or they
will decay to non-uniform or inhomogeneous black brane solution as
has been argued in \cite{Horowitz:2001cz, Horowitz:2002ym}. To
substantiate our claim, we can compute the free energy of the two
configurations and see whether the large black D$p$-branes has more
free energy than the small black D$p$-branes. The Helmholtz free
energy of the black D$p$-branes per unit $p$-brane volume is given
as, \bea\label{freeener} F &=& \frac{M-TS}{V_p}\nn &=&
\frac{2Q}{(7-p)g_s \sinh2\theta}\left[1+(7-p)\sinh^2\theta\right]
\eea where we have used the expressions for the ADM mass, the
entropy given in eq.\eqn{massentropy}, the temperature given in
eq.\eqn{temp} and the expression for the charge in eq.\eqn{charge}.
In the following we will look at the reduced free energy, which is
only relevant for the discussion when the charge and the temperature
are fixed, and is defined as, \be\label{redfree} \frac{(7-p)
g_s}{2Q} F \equiv \bar{F} = \frac{1+(7-p) \sinh^2\theta}
{\sinh2\theta} \ee We will first consider the case $p=4$ since in
this case we can compute the difference of reduced free energy
explicitly for all values of $C < C_{\rm max}$. For $p<4$, as we
will mention, we can compare the free energies of the two black
branes only for $C \ll C_{\rm max}$. From eq.\eqn{maxvalue}, we find
that for $p=4$, $C_{\rm max}$ has the value 1/2. So, for any
$C(\theta) < 1/2$, we have two values of $\theta$, namely,
$\theta_{\ell}$ (large $\theta$) corresponding to small black
D4-brane and $\theta_s$ (small $\theta$) corresponding to large
black D4-brane. The values of $\theta_{\ell,s}$ can be computed from
eq.\eqn{ctheta} as \bea\label{thetasoln} \sinh\theta_{\ell} &=&
\frac{1}{2C}\left(1 + \sqrt{1 - 4C^2}\right)\nn \sinh\theta_s &=&
\frac{1}{2C}\left(1 - \sqrt{1 - 4C^2}\right) \eea Substituting
\eqn{thetasoln} in \eqn{redfree} we find, \bea\label{diff}
\Delta\bar{F} &=& \bar{F}_{{\rm large\,D}4} - \bar{F}_{{\rm
small\,D}4}\nn &=&
\frac{\sqrt{2}(1-2C)}{4C}\left[\left(1+\sqrt{1-4C^2}\right)^{\frac{1}{2}}
- \left(1-\sqrt{1-4C^2}\right)^{\frac{1}{2}}\right] \eea So,
$\Delta\bar{F}$ is always positive as $C<1/2$. Thus for given charge
$Q$ and temperature $T$, we find that the large black D4-brane has
higher free energy than the small black D4-brane, consistent with
the specific heat analysis.

For $p<4$, it is difficult to solve the general $C(\theta)$ equation
(see eq.\eqn{ctheta})
\be\label{Ceqn}
\sinh\theta = C \cosh^{6-p}\theta
\ee
explicitly to find out $\theta_{\ell}$ and $\theta_s$. However,
the equation \eqn{Ceqn} can be solved when $C(\theta) \ll C_{\rm max}$.
In this case the small $\theta$ satisfies $\theta_s\ll 1$ while the large
$\theta$ satisfies $\theta_{\ell} \gg 1$.
So, for small $\theta$, we find, to the leading order (from eq.\eqn{Ceqn})
\be
\theta_s = C
\ee
while for large $\theta$ we find to the leading order,
\be
e^{\theta_{\ell}} = 2 C^{-\frac{1}{5-p}}
\ee
So, to the leading order
\bea
\bar{F}_{\rm large\,\,brane} &=& \frac{1}{2C}\nn
\bar{F}_{\rm small\,\, brane} &=& \frac{7-p}{2}
\eea
From here it is obvious that the large black brane has larger
free energy than the small black brane since $1/C \gg 1/C_{\rm max}$.
For example, if we take $1/C = 10/C_{\rm max}$,
then certainly large black brane has larger free energy. Once again,
this is consistent with the specific heat analysis.

Next we will look at the small black D$p$-brane
solutions which are classically stable for $0 < p \leq 4$ and
unstable for $p=5,6$. Small black branes even for $p \leq 4$ can
become quantum mechanically unstable. If we do not impose any
further conditions, all these black branes (including the
classically stable ones (for $ p\leq 4$) as well as the classically
unstable ones (for $p=5,6$))\footnote{Since black D5, D6 branes
are classically unstable even when they are small, there is a
possibility for them to decay to non-uniform (or inhomogeneous)
branes.} will Hawking radiate and if by this
process $\theta \to \infty$ such that the charge
remains fixed, the black brane will make a smooth transition to the
stable BPS D$p$-brane configuration (as is clear from
\eqn{blackbranesoln}, \eqn{functions}, \eqn{charge}). On the other
hand, if one of the brane directions is made compact, then the
fermions in the theory can be either periodic or antiperiodic along
the compact direction. For the periodic boundary conditions, the
black branes will Hawking radiate and eventually go over to the
stable BPS D$p$-brane solution as before. But, for the antiperiodic
boundary condition, the fundamental string wound along the compact
direction can become tachyonic. As argued in
ref.\cite{Adams:2005rb}, if the tachyon is localized it can trigger
a topology changing transition and the black D$p$-brane can decay
into either a static or a dynamical KK BON (for $0 < p \leq 4$) or
only to a dynamical KK BON (for D5 and D6 branes) by a closed string tachyon
condensation \cite{Horowitz:2005vp, Lu:2007bu}. However, in order
for the black brane to decay through the perturbative stringy
process of closed string tachyon condensation several conditions
have to be satisfied and we will outline them briefly here for
completeness, the details of which is given in ref.\cite{Lu:2007bu}.
First of all, note from \eqn{blackbranesoln} that, the size of the
compact circle (let us assume that $x^1$ is the compact direction)
varies from a finite value ($L\gg l_s$, where $l_s$ is the
fundamental string length) at infinity to zero at the singularity
($r=0$). So, in between the size of the circle becomes of the order
of $l_s$, and so the winding fundamental string along this direction
can become tachyonic. We will further assume that this happens on
the horizon\footnote{If the size of the circle is greater than $l_s$
on the horizon, there will be no closed string tachyon. However, as
the black D$p$-brane Hawking radiate, the horizon size gets reduced
and the size of the $x^1$-circle will also get reduced and
eventually, will attain the size $l_s$ and then the closed string
tachyon will appear \cite{Horowitz:2005vp}.} such that,
\be\label{condition1} L = l_s \cosh^{1/2}\theta \ee Since $L \gg
l_s$, eq.\eqn{condition1} implies that the parameter $\theta$ must
be very large or the size of the black brane is very small (this is
the classically stable region for $0 < p \leq 4$ as can be seen
from Figure 1). In that case the closed string tachyon will cause
the circle to pinch off and the black D$p$-brane will make a
transition to KK BON. The final KK BON configuration can be either
static or dynamical as we have argued in detail in \cite{Lu:2007bu}.
In order to have a static bubble the curvature on the horizon must
remain much smaller than the string scale, otherwise the
supergravity description will break down and the black D$p$-brane
will make a transition to open string modes (correspondence point)
\cite{Horowitz:1996nw}. Further, in order for the transition to
occur, the size of the horizon must match with the size of the
bubble, the charge of the black D$p$-brane must be equal to the flux
associated with the bubble and also the size of the $x^1$-circle at
infinity must match for the two configurations. We have seen in
\cite{Lu:2007bu} that all these conditions can be satisfied for the
black D$p$-branes with $0 < p\leq 4$ and the black D5 and D6
branes can only make transitions to dynamical bubbles.

We have thus seen how some simple observations on the black
D$p$-brane parameters make their stability analysis much easier. In
particular, we have shown that the charge, $Q$, and the Hawking
temperature, $T$ of the black D$p$-branes are such that the product
$QT^{7-p}$ has an upper bound for all $p \leq 5$, but black D6-brane
does not have such a bound. For $p \leq 4$, the maximum occurs at a
finite value of the parameter $\theta$, but for $p=5$, the maximum
occurs when $\theta \to \infty$. This implies that, for fixed
charge, there are two black D$p$-brane solution (for $p\leq 4$), one
small and the other large, at a given temperature below the maximum
value, whereas, for $p=5,6$, there is only one D$p$-brane solution.
By looking at the slope of the curve $C(\theta)$ vs. $\theta$
(depicted in Figure 1), which is proportional to the negative of the
specific heat of the black D$p$-branes, we found that, the large
black D$p$-branes are classically unstable, whereas, the small black
D$p$-branes are classically stable (for $0 < p\leq 4$) and D5 and
D6-branes are always classically unstable. The classically unstable
black branes have been argued to make a transition either to a
stable small black brane (for $0 < p \leq 4$) or to a non-uniform or
inhomogeneous stable brane configurations. Indeed, we have seen from
the computation of the free energies of the two black brane configurations
that the large black brane has larger free energy indicating that the
transition from the large black D-brane to the small black D-brane is
possible. However, when the
parameter $\theta$ becomes very large or the horizon size of the
black D$p$-brane becomes very small (this implies that the
temperature becomes very small for $p \leq 4$, but not for $p=5,6$),
both the classically stable D$p$-branes (for $0 < p \leq 4$) and
classically unstable D5, D6-branes can become quantum mechanically
unstable. So, the black branes (for all $p \leq 6$) will Hawking
radiate and will make a smooth transition to BPS D$p$-branes if we
do not impose any further conditions. But if one of the brane
directions is made compact (this can occur for $p>0$), then
depending on the periodicity or the antiperiodicity of the boundary
conditions on the fermions along the compact direction, the black
brane will make a transition either to a BPS D$p$ brane or to a KK
BON. The KK BON could be static or dynamical for $0 < p \leq 4$,
but for $p=5,6$, the KK BON will be dynamical.

\vspace{.5cm}

\section*{Acknowledgements:}

JXL acknowledges support by grants from the Chinese Academy of
Sciences, a grant from 973 Program with grant No: 2007CB815401 and
grants from the NSF of China with Grant No:10588503, 10535060 and
10975129.

\end{document}